\documentclass[%
preprint,
showpacs,preprintnumbers,
bibnotes,
 amsmath,amssymb,
 aps,
prstper,
longbibliography,
showkeys,
]{revtex4-1}

\usepackage{graphicx}
\usepackage{dcolumn}
\usepackage{bm}
\usepackage[english]{babel}

\usepackage{color}

\begin{document}

\preprint{Submitted to PRST PER}

\title{Assessing the quality of a student-generated question repository }


\author{Simon P. Bates}
 \email{simon.bates@ubc.ca}
 \affiliation{Department of Physics and Astronomy, University of British Columbia, Vancouver, Canada, V6T 1 Z1}
\author{Ross K. Galloway}%
\author{Jonathan Riise}
\author{Danny Homer}
\affiliation{
 Physics Education Research Group,\\
 School of Physics and Astronomy, University of Edinburgh, Edinburgh EH9 3JZ, United Kingdom
}


\date{\today}

\begin{abstract}
We present results from a study that categorizes and assesses the quality of questions and explanations authored by students, in question repositories produced as part of the summative assessment in introductory physics courses over the past two years.  Mapping question quality onto the levels in the cognitive domain of Bloom's taxonomy, we find that students produce questions of high quality. More than three-quarters of questions fall into categories beyond simple recall, in contrast to similar studies of student-authored content in different subject domains. Similarly, the quality of student-authored explanations for questions was also high, with approximately 60\% of all explanations classified as being of high or outstanding quality. Overall, 75\% of questions met combined quality criteria, which we hypothesize is due in part to the in-class scaffolding activities that we provided for students ahead of requiring them to author questions.

\end{abstract}

\maketitle


\section{\label{intro}Introduction}

One of the key features of the coming-of-age of modern information technologies (for example, within the so-called `Web2.0' movement) has been a shift from a single, authoritative content-owner, dispensing knowledge and information to those who consume it, towards a much more collaborative approach, with potentially large numbers of co-producers of content.  An often-cited example is the development of Wikipedia which, despite some concerns over the quality of some content, continues to be one of the ten most frequently accessed websites across the world \cite{wikipedia}, with a team of committed content authors well in excess of 300,000 and over 18 million occasional contributors \cite{wikipedians}. This article describes application of the same principle of co-production in the context of student-authored multiple choice assessment questions (MCQs) in introductory physics courses delivered at a large, research-intensive university in the UK. Specifically, we consider the quality of questions that students authored as part of their assessed coursework, and the explanations associated with these questions. 

It has been argued that there are specific and defined educational benefits from students being engaged in the co-creation as well as consumption of educational content \cite{Draper2009}. Cognitively, it can be far more challenging to have to create an assessment activity, for example a question complete with solution and explanation, than it is to simply answer one created by someone else.  It can require higher order skills far above simply `remembering' or `knowing' in a facile sense, a process which will be familiar to many faculty as they regularly take up the challenge of setting end-of-course assessments that meaningfully assess learning goals. Other authors have framed the benefits associated with student-generated content in terms of a participatory learning approach \cite{Bieber2005} designed to foster deep as opposed to surface learning. 
Earlier studies in psychology have shown that the act of question writing can be an effective study and learning technique, with question authors outperforming non-authoring students on subsequent tests, irrespective of whether they wrote essay-type or multiple choice questions \cite{Foos1989}.
Multiple choice questions are often viewed as quite limited in terms of their assessment potential, particularly when their use is driven by staff needs for greater efficiencies, rather than student needs for deeper understanding. However, it has been shown that they can support the process of formative assessment and feedback, as described by the `seven principles' of good practice identified by Nicol and Macfarlane-Dick \cite{Nicol2006}. A key finding from a related review \cite{Nicol2007}, which focused on effective e-assessment by design using MCQs, highlighted the importance of not just the questions themselves, but the context in which they were deployed within courses. A case study within this review presents student creation of assessment questions as a powerful articulation of Nicol and McFarlane-Dick's first principle, that students understand what is required for good performance in terms of goals, criteria and standards.

Skills in the cognitive domain of Bloom's taxonomy \cite{Bloom1956} provide a useful framework for categorizing question types and the activities associated with their production. Ascending the taxonomy levels (which are often represented diagrammatically as a pyramid structure) are descriptors  of: knowledge -- understanding -- application -- analysis -- synthesis -- evaluation. Anderson and Krathwohl \cite{AndersonKrathwohl2001} have suggested a simplification or revision to this structure that brackets the uppermost three levels of analysis--synthesis--evaluation together into a single compound category. A similar categorization, utilizing the same taxonomy, has been proposed as an aid to creating and refining the learning outcomes or goals associated with courses \cite{CWSEI}. In the work we report in this paper, our basic premise is that by requiring students to develop assessment content themselves, we are challenging them to operate at higher cognitive levels than they might otherwise do in the normal course of their studies.

Along with instructors at more than seven hundred institutions worldwide, we have been using the PeerWise online tool \cite{Denny2008} as the technology platform for these interventions. Developed in the Department of Computer Science at the University of Auckland, PeerWise is a freely-available, online tool to facilitate cohorts of students writing their own MCQs and answering and commenting on those of their peers. It incorporates much of the social functionality found in common popular websites, such as the ability to rate and comment on posts, and `follow' other contributors. At the time of writing, nearly 100,000 student registrants have contributed around 600,000 questions and approximately 12 million answers. In the last few years, several studies have emerged that have assessed the impact on student engagement and learning \cite{Denny2008b,Hakulinen2010a,Hakulinen2010b,Denny2011,Bottomley2011,Hooper2011,Luxton-Reilly2011,Paterson2011,Sykes2011,Paterson2012} including our own initial study \cite{Bates2012} of piloting PeerWise as a summatively-assessed component in introductory physics courses. Most report positive student engagement with and attitudes towards the system and several evidence a correlation between usage of the system and end-of-course outcomes (e.g.\ the final exam grade) \cite{Denny2008b,Hakulinen2010b,Luxton-Reilly2011,Sykes2011,Bates2012} that impacts on students of all abilities in the course. 

Far fewer studies have featured a discussion of the quality of questions authored by students. Hakulinen \cite{Hakulinen2010a,Hakulinen2010b} and Korhonen \cite{Hakulinen2010a} used expert ratings to investigate the reliability of the students' own quality ratings.  They also explored automated approaches and expert sampling to classify the quality of questions on a binary basis (`good' vs.\ `bad'), finding that the majority of student contributions were of `good' questions. Similarly, an investigation by Purchase {\em et al.} \cite{Purchase2010} of characteristics such as topic coverage, holistic question quality, difficulty, and indexing of questions submitted by a first year programming class found that the question repository was of generally a high standard. Bottomley and Denny \cite{Bottomley2011} report a study of a cohort of 107 second year biomedical science students. Over 90\% of contributed questions were found to be correct and of those incorrect questions, approximately half were identified as such by students who answered and/or commented on the question. The majority of questions contributed by students were classified at the lowest taxonomic level (`recall' / `remembering') with less than 10\% above level 2 (`understanding') of Bloom's taxonomy.  The authors state that this is to be expected, since for these students this was likely the first time they had been challenged to write their own questions. This finding is to be compared to similar studies that have used the same mapping procedure to categorize instructor-authored questions onto the levels of Bloom's taxonomy. One in particular presents surprising findings: a recent study \cite{Momsen2010} examining 9,713 assessment items submitted by 50 instructors of introductory biology courses in the United States reported that 93\% of the questions were at the lowest two levels of the revised Bloom's taxonomy. Zheng {\it et al.} \cite{Zheng2008} have applied the same procedure to provide evidence that the questions in MCAT examinations are strong from this perspective, but find similar high proportions of questions at the lowest levels of Bloom's taxonomy in other university examinations. 

Given the widespread use of the PeerWise system, yet the paucity of reported studies as to the quality of contributed questions, it is both timely and necessary to address this issue. This paper reports a comprehensive evaluation of the question and explanation quality in student-authored questions across two separate, consecutive introductory physics courses delivered over two successive academic years at the University of Edinburgh.  The paper is organized as follows: in the next section, we report brief details of the educational context of the courses and the specific details of PeerWise implementation in the courses, together with the post-hoc analysis procedure. We then present results of the question quality, mapping onto the levels of Bloom's taxonomy, and the quality of explanations using a classification rubric of our own devising. We briefly present initial findings in terms of analysis of numbers of student responses as a function of question quality, before discussing our results more broadly. We end with some conclusions and suggestions for further research. 

\section{\label{method}Methodology}

The educational context for this intervention is a pair of consecutive introductory physics courses in the first year of the physics program at the University of Edinburgh. Physics 1A is a first course in classical mechanics and statics, covering kinematics, Newton's Laws, energy, momentum, rotational motion, and oscillations. Physics 1B is a `showcase' course in modern physics, covering broad topics at an introductory level including quantum mechanics, thermal, nuclear and particle physics. Both are taken by a broad and diverse student cohort in terms of prior study and future aspirations. Approximately half of the cohort are studying towards a Physics degree. The remainder are taking the courses as an elective subject; these students are as equally-qualified (in terms of high school physics and mathematics grades) as the Physics majors. Approximately three-quarters of the cohort are male, and the vast majority of all students are aged between 17 and 19. More than 95\% of the first semester Physics 1A cohort go on to take Physics 1B. Both courses have employed a variety of research-based and interactive engagement strategies, including extensive use of clickers, and studio-based workshop teaching \cite{Bates2005} (similar to the TEAL \cite{TEAL} model). In Physics 1A, we have employed the FCI \cite{FCI}
at the start and end of the course to gauge both incoming cohort ability and effectiveness of instruction. Typical FCI results, which are relatively stable over several years, are a pre-instruction cohort average of around 65\% and a post-instruction normalized gain value of around 0.5--0.6. We have previously reported results \cite{CLASS_us} of pre-post CLASS \cite{CLASS} scores for the cohort, with the most significant finding being the high on-entry overall agreement with expert views (around 69-72\% over the past 4 years). 

This study reports on data from the incorporation of PeerWise  activities into the summative assessment of both the Physics 1A and 1B courses, over two successive academic sessions (2010-11 and 2011-12 academic years, hereafter referred to as 2010 and 2011 data for simplicity). The first of these two years was a pilot implementation: one weekly assessment task in each of Physics 1A and 1B was replaced by a PeerWise activity in which students were required to contribute, as a minimum, one original question that they authored, answer five others, and comment and rate on a further three questions. In Physics 1A 2010, the PeerWise activity was launched in week 5 of the semester, with the assessment deadline one week later. In Physics 1B 2010, an identical assessment requirement was set in week 4 of the semester, with a due date at the end of teaching in week 11. In each case, the PeerWise assessment contributed approximately 3\% of the summative assessment grade for the course, with the scoring system built into the PeerWise system serving as the basis for allocation of assessment marks (for further details see Bates {\em et al.} \cite{Bates2012}). In the 2011 deployment in Physics 1A, we replaced three weekly assessments with PeerWise assessment activities, and in Physics 1B 2011 just one weekly assessment activity. As previously, students were required to contribute, as a minimum, one authored question, to answer five others, and to rate and comment on a further three in each activity. 

For both years of deployment, a significant component (90 minutes) of one of the class sessions was devoted to preparatory activities ahead of the first PeerWise assessment activity of the year. These sessions, deliberately designed to help scaffold the process of writing questions of high quality, comprised four elements: 

\begin{itemize}
\item A content-neutral quiz \cite{PhilRace} that taught the language of MCQs (stem, options, key, distractors) and demonstrated how poorly written questions sometimes test nothing but language skills.  
\item A magazine-style self-diagnosis quiz to help students to explore their beliefs about thinking and guide them toward learning orientation and away from performance orientation.
\item A question template introducing students to the notion of challenging themselves to write questions at a level just beyond their current understanding, aligned with the notion of operating in Vygotsky's `Zone of Proximal Development' \cite{Chaiklin2003}. This simplified constructivist model, along with information about common misconceptions and errors, encouraged the students to author questions of high cognitive value.
\item A good quality example question, based on the template, which set a very high bar for expected creativity and complexity.
\end{itemize} 

We devoted approximately 90 minutes of class time to covering these four elements, with a final activity in which students worked in groups of five or six to collectively author a question using our template. These group-authored questions were then uploaded to the online system to seed the database prior to setting the assessment task. All of our scaffolding materials are freely available online \cite{scaffolding}. 

Analysis of question and explanation quality was conducted post-hoc. Two of the authors (JR and DH), undertaking final year undergraduate honors projects and working in collaboration with the faculty member authors (SPB and RKG), devised and agreed on a series of three classifications to determine question quality. The first of these classified the cognitive level of the question based on the levels in Anderson and Krathwohl's revised version \cite{AndersonKrathwohl2001} of Bloom's taxonomy \cite{Bloom1956}, illustrated and summarized in Table~\ref{tab:taxonomy}. When undertaking this classification, the question itself, the question setter's provided solution, and subsequent comments in the question's comments thread were all visible to the rater: this wide range of information (and, in particular, the availability of the solution) made assigning a classification for cognitive level more straightforward than if the question alone had been available.
\begin{table}[htb]
\caption{\label{tab:taxonomy}
 Categorization levels and explanations for the cognitive domain of Bloom's taxonomy}
\begin{ruledtabular}
\begin{tabular}{cll}
\textrm{Level}&
\textrm{Identifier}&
\textrm{Explanation and interpretation}\\
\colrule
1 & Remember & Factual recall, knowledge, trivial `plugging in' of numbers. \\
2 & Understand & Basic understanding, no calculation necessary. \\
3 & Apply & Implement, calculate or determine. Single topic \\
&& calculation or exercise involving application of knowledge. \\
4 & Analyze & Typically multi-step problem; requires identification \\
&&of problem-solving strategy before executing. \\
5 & Evaluate & Compare and assess various option possibilities; \\
& & often qualitative and conceptual questions. \\
6 & Create & Synthesis of ideas and topics from multiple course \\
& & topics to create significantly challenging problem. \\
\end{tabular}
\end{ruledtabular}
\end{table}

The second scale classified the quality of the explanation and solution associated with each question (which student contributors are required to provide at the time of authoring the question). This categorization scheme is illustrated in Table~\ref{tab:explanation}.

\begin{table}[htb]
\caption{\label{tab:explanation}
 Categorization levels for explanation of solution to questions}
\begin{ruledtabular}
\begin{tabular}{cll}
\textrm{Level}&
\textrm{Identifier}&
\textrm{Description}\\
\colrule
0 & Missing & No explanation provided or explanation incoherent.\\
1 & Inadequate & Wrong reasoning and/or answer. \\
& & Solution may be trivial, flippant or unhelpful.\\
2 & Minimal & Correct answer but with insufficient explanation or justification.\\
&& Some aspects may be unclear or incorrect.  \\
3 & Good & Clear and sufficiently detailed exposition of both correct \\
&& method and answer. \\
4 & Excellent &Thorough description of relevant physics and solution strategy. \\
&& Contains remarks on plausibility of answer  and/or other distractors. \\
&& Beyond normal expectations for a correct solution.
\end{tabular}
\end{ruledtabular}
\end{table}

Finally, overall criteria were devised to determine whether or not a question could be judged to be a high quality question. These criteria included required characteristics based on cognitive level of the question (above factual recall, i.e.\ Bloom's level 2 or higher) and explanation quality (`minimal' or higher), together with other measures. A question was classified as high quality if it met all of the criteria outlined in Table~\ref{tab:criteria}. Requiring at least two plausible distractors eliminates `yes/no' or `true/false' questions from being classified as high quality. In terms of identifying correctness and originality, the caveats `not obviously' (plagiarized) and `most likely' (correct) were added as it was not practical to cross-reference all questions with all those publicly available (as end-of-chapter problems or on the web), nor to work through numerical solutions to all the problems contained within sampled questions. However, for a representative sample of questions, we pasted the question stem text into a search engine to check against material openly available on the internet and confirm the originality of the questions. We found no evidence of plagiarism from internet sources. Furthermore, the innovative contexts for many questions (e.g.\ see examples in the Supplementary Material) are clearly different in style to those usually found as end-of-chapter problems.

\begin{table}[htb]
\caption{\label{tab:criteria}
 Criteria used to define high quality questions}
\begin{ruledtabular}
\begin{tabular}{ll}
\textrm{Measure}&
\textrm{Criteria details}\\
\colrule
Taxonomy category & At least level 2 or higher (`understand' or above). \\
Explanation category & At least level 2 or higher (`minimal' or better). \\
Clearly worded question & Unambiguous vs.\  unclear (binary measure). \\
Distractors & At least 2 feasible and plausible distractors. \\
Correctness & `Most likely' correct (binary measure). \\
Plagiarism & `Not obviously' plagiarized (binary measure).  \\

\end{tabular}
\end{ruledtabular}
\end{table}

For categorization of question cognitive level and explanation quality, inter-rater reliability tests were undertaken to ensure consistency between the two coders. Each of the coders categorized a sample of questions from different course repositories for the cognitive level and explanation quality, before exchanging sample question sets and classifying each other's. The inter-rater reliability was subsequently determined by calculating Cohen's Kappa \cite{Kappa}. Although there are several different methodologies to determine inter-rater reliability, there are a number of advantages to Cohen's Kappa: it naturally accounts for expected purely coincidental agreement between raters, and handles discretized rating schemes well. It is usually  considered a robust and rather conservative estimate of the inter-rater reliability. An initial sample of 35 questions produced moderately good inter-rater reliability. Following discussion between the raters and with the faculty member authors, a further 22 questions were sampled and coded from each repository. This yielded very good inter-rater reliability between all the contributors (over 90\% for both question level and explanation quality). 

In other projects, we have established that for undergraduate students working within our group on PER projects, the initial inter-rater reliability between student and expert (faculty) rater is over 70\%. Following calibration and discussion of an initial set of ratings of questions, this level of agreement rises to over 90\%. Furthermore, it is sustained at this level several weeks later without any further intervention or calibration in the mean time. This provides strong evidence that these students learn how to do this classification well and can undertake it very reliably. 

A representative sample of the four distinct question repositories were coded, 602 questions in total. These were divided across the four courses as illustrated in Table~\ref{tab:numbers}. The absolute values give the actual number of questions coded from each repository, with the percentages illustrating what fraction of the total repository those numbers represent.

\begin{table}[htb]
\caption{\label{tab:numbers}
Number and percentages of question repository sampling across all courses}
\begin{ruledtabular}
\begin{tabular}{lcc}
\textrm{Course}&
\textrm{2010}&
\textrm{2011}\\
\colrule
Physics 1A & 150 (42\%) & 200 (24\%) \\
Physics 1B & 179 (52\%) & 73 (46\%) \\
\end{tabular}
\end{ruledtabular}
\end{table}

Tests of statistical significance to identify differences in the distributions of cognitive level or explanation quality of coded questions, between courses or between years, were calculated by performing $\chi$-squared tests on actual and expected distributions.  We assigned statistical significance to test statistics with $p<0.05$.

\section{\label{results}Results}

In this section we present the results of classifying questions from the various course repositories. The Supplementary Material for this paper presents case-studies of a small number of student-authored questions, where we detail the rationale for why the questions and explanations were classified in the selected categories, and present brief details of the discussions that took place within the student cohort around these particular questions. 

\subsection{Cognitive level of questions}

\begin{figure*}
\includegraphics[width=4.5 in]{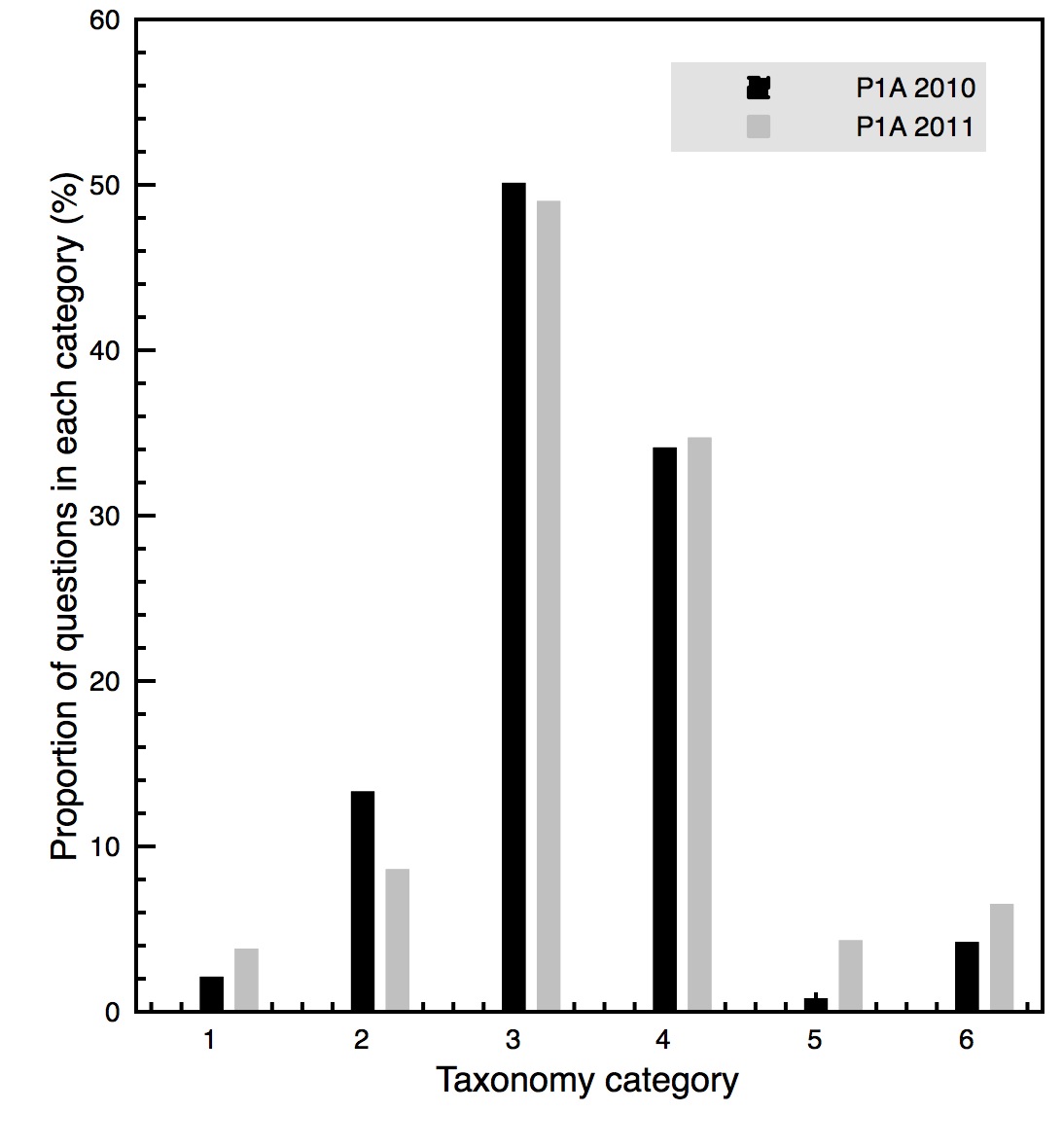}
\caption{\label{fig:tax1A}Proportion of questions in each taxonomic category, for Physics 1A question repositories for 2010 ($N$=150, dark bars) and 2011 ($N$=200, light bars).}
\end{figure*}

Categorizations of the cognitive level of questions sampled from repositories of one course (Physics 1A) over two successive years are shown in Figure \ref{fig:tax1A}, based on the taxonomy level descriptions from Table \ref{tab:taxonomy}. For both years, there is only a small proportion of questions in the lowest taxonomy category (less than 5\% for both years) and the majority of student questions are categorized as those requiring application or analysis, usually in the form of a quantitative problem to be solved in one or multiple stages, respectively. A 
$\chi$-squared
 test indicates that there is no statistically significant difference ($p$=0.27) between the distributions from the two years.

\begin{figure*}
\includegraphics[width=4.5 in]{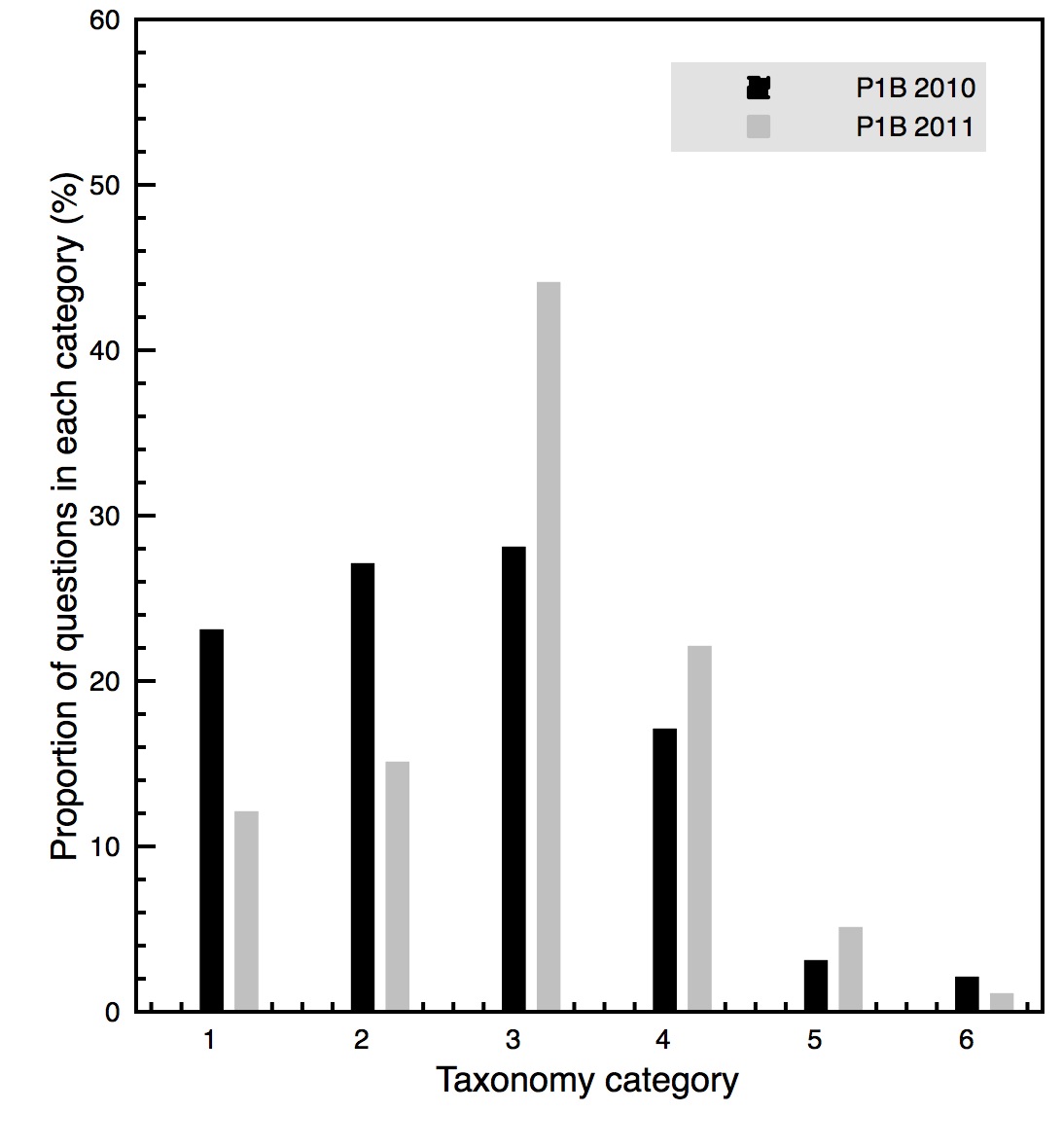}
\caption{\label{fig:taxcomb}Proportion of questions in each taxonomic category, for Physics 1B question repositories for 2010 ($N$=179, dark bars) and 2011 ($N$=73, light bars).}
\end{figure*}

 Figure \ref{fig:taxcomb} presents the equivalent data for the set of questions sampled from the Physics 1B repositories over two successive years. Once 
 again, we find a distribution spanning all taxonomic categories, but with some interesting differences to the distribution in Figure \ref{fig:tax1A}. Firstly, there is a difference in the shape of the profile between the two successive courses in both of the two years under study. Recall that the cohorts for Physics 1A and 1B in any given year are essentially the same group of students. This suggests that there is an influence played by the nature of course material on the types of questions students create. In the Physics 1A course (covering introductory topics in mechanics and oscillations) there is a different profile to the 1B course (designed as `grand tour' of modern physics, covering more material in somewhat lesser detail). This is not altogether surprising since it is natural to assume that the particular subject matter and design of the course will have some bearing on the nature of the student-authored questions.

 In contrast to the results for Physics 1A, Figure  \ref{fig:taxcomb} exhibits a statistically significant difference ($p$=0.022) between sampled 1B questions from subsequent years. In particular, there is a  smaller fraction of questions in the lower taxonomic levels in the 2011 data. This may be a result of the different implementation strategies in the two years. In 2010, a single PeerWise activity was introduced into Physics 1A. In 2011, three separate activities were undertaken in 1A. Thus, the 2011 1B cohort has had greater experience and practice in authoring questions. An interesting direction for future study would be to explicitly test this `quality improves with practice' hypothesis. 


\subsection{Explanation quality}

A similar analysis was undertaken for the quality of student-authored explanations associated with each question, using the classification rubric shown in Table \ref{tab:explanation}. The PeerWise system does not {\em require} the explanation field to be completed prior to submitting the question into the repository. However we made it clear to students that the ability to articulate the solution strategy, together with the ability to explain why incorrect answers were wrong, was a required and important part of developing a question. 

\begin{figure*}
\includegraphics[width=4.5 in]{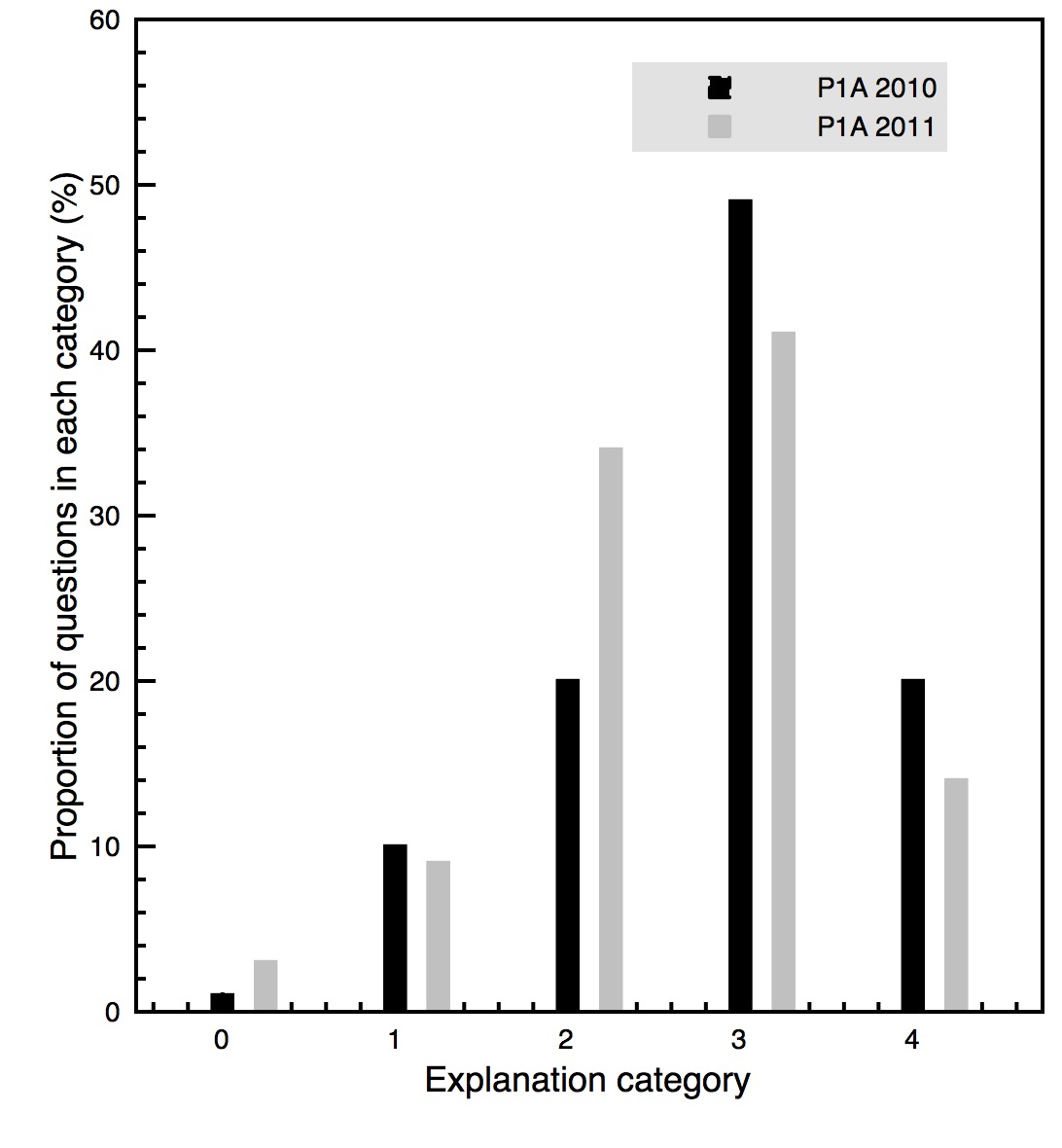}
\caption{\label{fig:exp1A}Proportion of questions in each explanation category, for Physics 1A question repositories for 2010 ($N$=150, dark bars) and 2011 ($N$=200, light bars).}
\end{figure*}

Figure \ref{fig:exp1A} shows data from questions sampled from the 2010 and 2011 Physics 1A question repositories. The figure shows that in over 95\% of cases students did construct an explanation of some kind, and in the vast majority of cases these were of good or excellent quality (approximately two-thirds of the questions sampled were in the `good' or `excellent' categories for each of the two years). What is particularly impressive is the proportion of questions in the uppermost explanation category, those which went far beyond what might have been ordinarily expected from a student solution to a problem. Many of the explanations in this  category demonstrated important developmental skills within the discipline: sense-making of the answer; appealing to special cases; multiple routes to solve the same problem; and articulation of commonly-held alternate conceptions when discussing distractors. 

A $\chi$-squared test revealed a statistically significant difference between the distributions of the explanation categories for the Physics 1A data for 2010 and 2011 (shown in Figure \ref{fig:exp1A}) with $p$=0.022. The sampled explanations from the 2011 cohort appear to be of slightly lower quality overall compared to 2010 data. With hindsight, this pair of distributions confirms our experiences as instructors that the three separate PeerWise assessments within a single semester, when viewed `in the round' with all the other course assessment tasks, were probably too many, evidenced by a noticeable drop in proportion of students engaging with the third assessment in 2011 (data not presented here). Notwithstanding, the overall profile of both distributions is striking, with very few questions lacking meaningful explanation. The modal classification was in the `good' category (between 40 and 50\% of all questions in each course sample) and a non-negligible fraction were in the highest category in both cases. 

\begin{figure*}
\includegraphics[width=4.5 in]{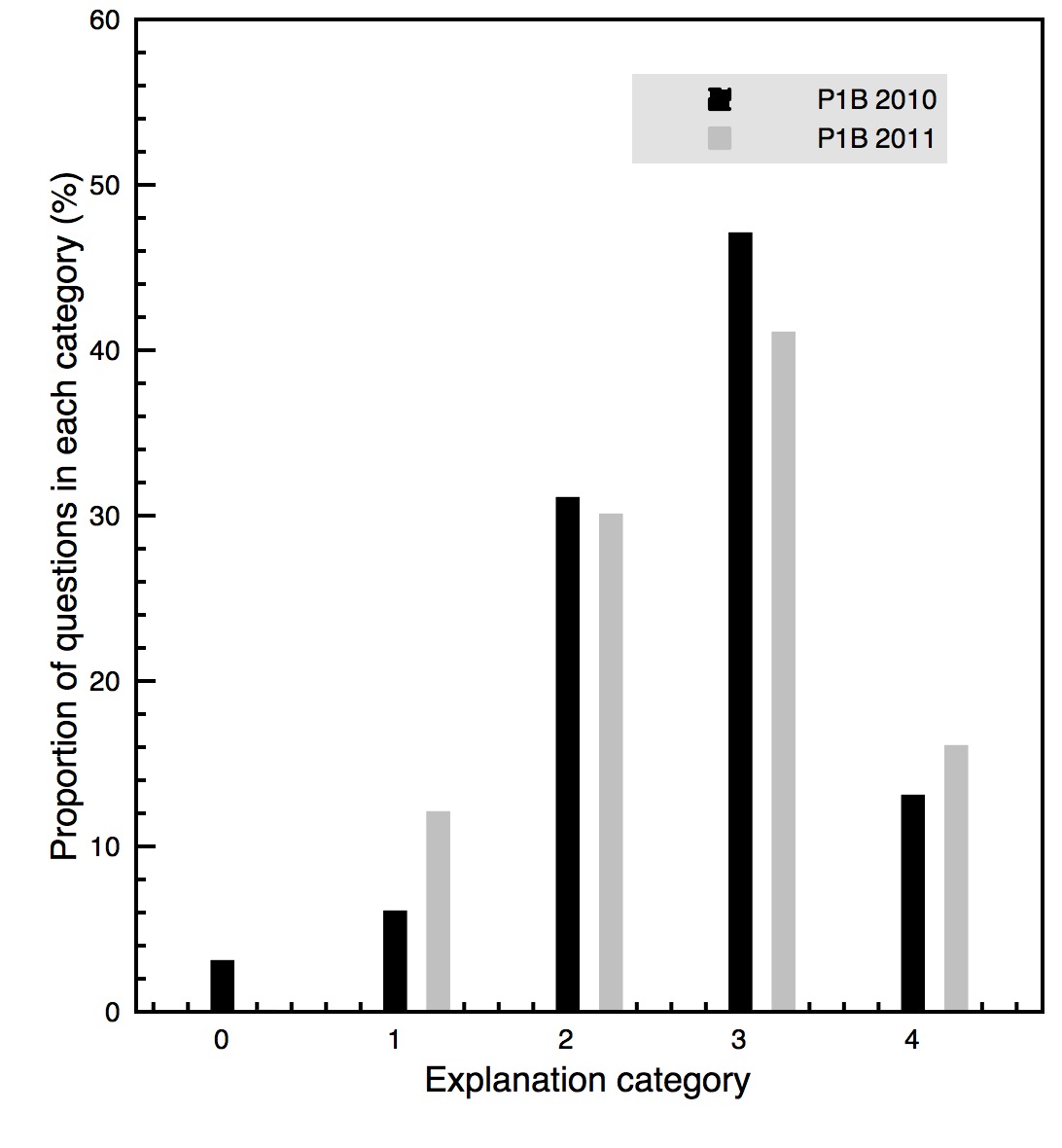}
\caption{\label{fig:expcomb}Proportion of questions in each explanation category, for Physics 1B question repositories for 2010 ($N$=179, dark bars) and 2011 ($N$=73, light bars).}
\end{figure*}

Equivalent data for the explanations associated with questions sampled from the 2010 and 2011 Physics 1B repositories are shown in Figure \ref{fig:expcomb}. Here, there is no statistically significant difference ($p$=0.66) between samples for two successive years, and the same overall pattern in the profile of explanation quality detailed above for the 1A questions is once again clearly apparent.


\subsection{Overall question quality}

To be classified overall as a high quality question, a submission was required to meet \emph{all} of the criteria presented in Table \ref{tab:criteria}. These include minimum requirements for the classification of cognitive level and explanation quality (at least level 2 or higher in both cases: `understanding' or above in terms of cognitive level, and a `minimal' level of explanation or above). In addition, these criteria also included further quality requirements pertaining to question clarity, plausibility of distractors, originality and correctness. 

Combining all sampled questions together ($N=602$), our classification yielded that overall 453 questions (75\%) met all the criteria outlined in Table \ref{tab:criteria}. In terms of individual criteria failure rates, our analysis showed that:

\begin{itemize}
\item On grounds of clarity, only 5\% of questions were rejected on the basis that the question statement was unclear, ambiguous or irrelevant;
\item In terms of the number of distractors, 80\% of questions had at least two plausible distractors;
\item 10\% of questions were rejected on the basis of having insufficient explanation;
\item 10\% of questions were rejected on the basis of too low a taxonomy classification;
\item Only 1\% of questions were identified as being obviously plagiarized (and in most cases, these were highly derivative of questions already in the repository, or of problems elsewhere in the course materials); 
\item Only 5\% of questions were rejected because they were identified as incorrect or seriously flawed in some way (and in more than half of these instances, the error/mistake had been identified by other students).
\end{itemize}

\subsection{Student answer patterns as a function of question quality}

An extensive treatment of student behavior in terms of question answering and the related comments/discussions is beyond the scope of this paper. However, we do present data to address the question ``Do students answer easier questions more frequently, and if so by how much?'' One might imagine a situation where students fulfill minimum assessment requirements by strategically targeting `easy' questions, thus putting in the minimum effort. 

\begin{figure*}
\includegraphics[width=4.5 in]{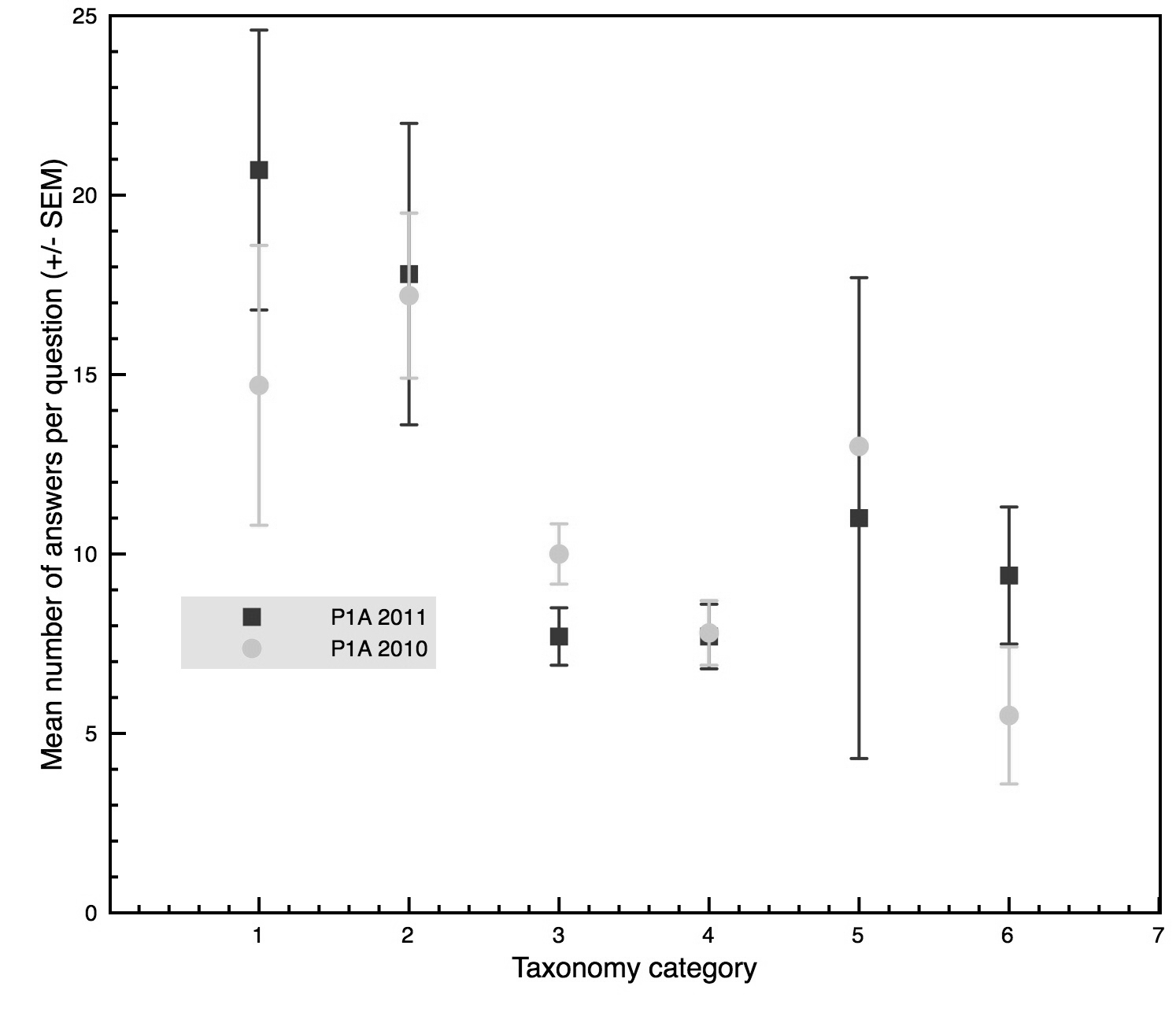}
\caption{\label{fig:Qvtax}Mean number of student responses per question as a function of question taxonomy, for Physics 1A course repositories for the 2010 and 2011 cohorts. Error bars denote the standard error on the mean. Note that for P1A 2010, there was only one category 5 question in the sample.}
\end{figure*}

For the 602 questions evaluated across the four course offerings, we have calculated the mean number of answers per question as a function of question taxonomic category. Representative data is shown for Physics 1A 2010 and 2011 in Figure \ref{fig:Qvtax}. With a relatively small number of distinct course datasets (4), we must, of course, be cautious about drawing too many conclusions. Nonetheless, the data suggest that questions in lower taxonomic categories tend to be answered more, but only by a relatively small factor. For example, in questions sampled from the Physics 1A 2011 repository, category 1 and 2 questions are answered on average 21(4) and 18(4) times, respectively, where the values in parentheses give the standard error on the mean. For questions in higher categories, the mean number of answers per question ranges from 8(1) to 11(7), though for smaller numbers of questions in higher categories the statistics can be somewhat distorted by one or two questions that tend to be extremely popular. The two different courses (Physics 1A and 1B) show broadly similar behavior: that is, questions in higher taxonomic categories tend to attract fewer answers per question. In summary, as illustrated by Figure \ref{fig:Qvtax}, there is evidence for reasonable consistency in patterns of answering in the same course over successive years: students answer the lower category questions more frequently, but do also answer a substantial number of the longer, more involved and more challenging higher category questions.

\section{\label{disc}Discussion}

The quality of the student-authored questions examined in this study, in terms of their categorization onto corresponding skills in the cognitive domain of Bloom's taxonomy, is significantly different to that reported previously in different subject domains. Bottomley and Denny \cite{Bottomley2011} report that 90\% of question items authored by biology undergraduate students lie in the lowest two taxonomy levels, with more than half in the lowest category. In contrast, we report a broad distribution across all taxonomic levels,  with a majority of the questions in the middle categories of `application' and `analysis'. Categorization is always a somewhat subjective activity: we have attempted to ensure that our process is as robust as possible with appropriate inter-rater reliability checks. However, it is certainly still possible that what one person might interpret as being appropriately categorized as `analysis' may be better classified as `evaluation' by someone else. That said, it is rather more straightforward, with a knowledge of the course material covered, to determine those questions that really are in the lowest category of factual recall rather than belonging higher up the classification. This is one reason why we have adopted the minimum criterion of taxonomy classification to be `understand' or higher for the question to be potentially judged as a high quality one: we may debate a question categorization at the highest levels, but in our experience everyone can consistently recognize one at the lowest. 

There are also similarities to be drawn with the study reported by Bottomley and Denny \cite{Bottomley2011}. They too find that over 90\% of questions are accompanied by an adequate explanation or better (rated on a four point scale, rather than our five point one). Likewise, they find similarly high proportions of correct solutions and not-obviously-plagiarized material as we report here. It is our hypothesis that the higher quality of student-authored questions found in the present study is connected to the introductory exercises and scaffolding activities that we provided to students ahead of the first PeerWise assessment task. These not only serve to set the bar at a high level in terms of expected contributions (by provision of a high quality worked example), but also provide support for (many, but by no means all) students to extend themselves beyond what they currently `know', thus challenging their own understanding. Ideally, to test this hypothesis we would seek to set up a controlled experiment that contrasts question quality from a control and two intervention groups (no PeerWise activity, PeerWise with no scaffolding, and PeerWise with scaffolding). We have not yet gathered such data, but do note that previously reported studies of question quality do not describe significant scaffolding or introductory activities. (Previous implementations of PeerWise which do report the use of scaffolding \cite{Paterson2011,Sykes2011} found enhanced levels of student engagement but did not investigate the quality of contributed questions.) Analysis of replication studies in other disciplines and institutions, utilizing similar scaffolding materials modified appropriately for local contexts, is under way and will be reported elsewhere. 

Our results indicate that not only do students produce, on the whole, very good questions, but also the appropriately detailed and useful explanations to accompany them. It may be the case that having ownership of a question encourages students to create more detailed commentary, as does the responsibility of contributing to a resource that will benefit their peers, plus the positive feedback received through comments and `badges' within the system. Informal consultations with students suggested that the median time to create and refine a question plus develop a solution was between 1 and 2 hours. Though there is considerable variation in how much time students spend engaged with the system, this seems an appropriate time investment for the summative assessment credit (typically 2-3\% of course grade) associated with each PeerWise task. 

We find differences in the distribution of questions across the taxonomic levels for different courses, but far less difference between different years of the same course (even though in the case of Physics 1A the use of PeerWise was increased from one to three activities in 2011). This is perhaps not surprising as the particular course content material will have a bearing on the types of questions students author. Physics 1A is a first course in classical mechanics, with rich contextualization possibilities from everyday scenarios. Physics 1B is a `grand tour' course of the fundamentals of modern physics, with a greater range of topics covered (thus in somewhat less depth); there are fewer obvious avenues for real-world contextualization of questions in this course as compared to Physics 1A. Despite these differences, the same broad conclusion of students being very capable of producing high quality questions (and explanations) holds. 

Our results suggest that questions that are classified in the lowest two taxonomic categories tend to be answered more frequently, despite there generally being fewer of them. One possible explanation for this observation is the time required to answer each type of question: a category 1 or 2 question could be solved by a few seconds' careful thought, enabling students to make rapid progress through them, whereas higher category questions might require tens of minutes of calculation and problem solving.  The difference in mean number of answers per question as a function of taxonomic category is around a factor of 2, which is somewhat smaller than might be expected if a large fraction of the cohort are being tactical and trying to answer the most straightforward questions for `easy marks'. This factor is also likely to be heavily influenced by the particular requirements for an assessed task: insisting that students answer a large number of questions as a minimum requirement is likely to lead to more tactical choice of questions to answer. Likewise, requiring students to set an unreasonable number of questions in proportion to the time/effort they have available is highly likely to result in questions of lower quality. Thus, we would suggest that the context in which the system is used in a course, together with the material and support provided to help students write questions, are both important factors that have a bearing on question quality.

\section{\label{conc}Conclusions and Future Work}

We have classified student-authored questions produced as part of the summative assessment for four introductory physics courses (two semester-long courses, over two successive academic sessions) according to cognitive level and quality of explanation. We find that these first year students are capable of producing very high quality questions and explanations. On the basis of minimum thresholds for cognitive level and quality of explanation, together with other question-specific criteria, we find that  75\% of the questions can be classified as being of high quality. Questions meeting these criteria are clear, correct, require more than simple factual recall to answer, and possess a correct solution and plausible distractors. In particular, a substantial fraction of the questions constitute true problems (as opposed to simple exercises). A significant difference between our implementation of PeerWise and other reported studies examining contributed question quality is the provision of support and scaffolding materials prior to the start of the assessment activity. 

Our previous work with PeerWise has demonstrated that incorporation into the summative assessment strategy for a course can lead to good engagement and enhanced learning \cite{Bates2012}, confirmed by similar studies in other disciplines \cite{Denny2008b,Hakulinen2010b,Luxton-Reilly2011,Sykes2011}. The present work complements these studies, indicating that students are capable of producing high quality questions and detailed explanations. These findings, coupled with the efficiency associated with student assessment being largely done by their peers, suggest that this instructional methodology can become part of the `standard toolkit' of student-centered course design for undergraduate physics. 

This remains a fruitful area for on-going research, including replication studies at different institutions and in different subject contexts, which we will report elsewhere. Further detailed analysis is underway of the student comments, and more broadly a learning analytics / student network analysis to understand interactions between question authors and answerers. Finally, given that the proportion of students taking introductory physics courses far exceeds those going on to study for a physics degree, we are investigating the incorporation of such strategies into classes composed entirely of non-majors. Our work with PeerWise in undergraduate classes suggests that students are substantially more creative than we might have previously given them credit for, and this creativity might be usefully harnessed in meaningfully developing core skills (such as problem solving) within the discipline. 

\begin{acknowledgments}
This work has been partly funded under a grant from the Joint Information Systems Committee (JISC), under their Assessment and Feedback strand. Significant contributions to the project have been made by other members of the Physics Education Research Group at the University of Edinburgh, including Judy Hardy, Karon McBride and Alison Kay. 

\end{acknowledgments}

\bibliography{refs}

\end{document}